\documentclass[%
 reprint,
superscriptaddress,
showkeys
amsmath,amssymb,
]{revtex4-1}

\usepackage{graphicx}

\begin{document}

\title{Superdiffusive motion of membrane-targeting C2 domains}
  
\author{Grace Campagnola}
\affiliation{Department of Biochemistry and Molecular Biology, Colorado State University, Fort Collins, CO 80523, USA}

\author{Kanti Nepal}
\affiliation{School of Biomedical Engineering, Colorado State University, Fort Collins, CO 80523, USA}

\author{Bryce W. Schroder}
\affiliation{School of Biomedical Engineering, Colorado State University, Fort Collins, CO 80523, USA}

\author{Olve B. Peersen}
\affiliation{Department of Biochemistry and Molecular Biology, Colorado State University, Fort Collins, CO 80523, USA}

\author{Diego Krapf}
\email[E-mail: ]{krapf@engr.colostate.edu}
\affiliation{School of Biomedical Engineering, Colorado State University, Fort Collins, CO 80523, USA}
\affiliation{Department of Electrical and Computer Engineering, Colorado State University, Fort Collins, CO 80523, USA}

\date{\today}


\begin{abstract}
Membrane targeting domains play crucial roles in the association of signalling molecules to the plasma membrane. 
For most peripheral proteins, the protein-to-membrane interaction is transient. After proteins dissociate 
from the membrane they have been observed to rebind following brief excursions in the bulk solution. 
Such membrane hops can have 
broad implications for the efficiency of reactions on membranes. 
We study the diffusion of membrane-targeting C2 domains using single-molecule tracking in supported lipid bilayers. 
The ensemble-averaged mean square displacement (MSD)  
exhibits superdiffusive behaviour. 
However, traditional time-averaged MSD analysis of individual trajectories remains linear 
and it does not reveal superdiffusion. Our observations are explained in terms of bulk excursions 
that introduce jumps with a heavy-tail distribution. 
These hopping events allow 
proteins to explore large areas in a short time. 
The experimental results are shown to be consistent with 
analytical models of bulk-mediated diffusion and numerical simulations. 
\end{abstract}

%%%%%%%%%%%%%%%%%%%%%%%%%%%%%%%%%%%%%%%%%%%%%%%%%%%%%%%%%%%%%%%%
\maketitle

A myriad of signalling proteins are recruited to specific cell membranes via phospholipid-binding domains 
\cite{Hurley2006805,lemmon2008membrane}. 
These molecules dock to the surface of specific lipid membranes and undergo two-dimensional diffusion in search of a target. Once the target is located, 
many proteins either activate or suppress a downstream signalling pathway for various physiological and pathological processes. 
Examples of membrane-targeting domains include pleckstrin homology (PH) \cite{lemmon2000signal} 
and C2 \cite{Cho2006838}, which have been identified in hundreds of 
human signalling molecules as well as in eukaryotic species as diverse as fungi and flies \cite{letunic2012smart}. 
PH domains bind specifically to phosphoinositides while C2 domains bind a variety of membranes, 
and a subset of C2 domains only bind membranes in the presence of calcium and play key 
roles in signalling pathways. 
The association to lipid membranes often takes place in response to different extracellular and intracellular stimuli, 
but typically the residence on the membrane surface is only temporary. The transient nature 
of peripheral protein-membrane interactions enables a tight temporal regulation of signal transduction. Further, 
membrane dissociation has also broad implications on the search for the target substrate, but this process is less understood.

Recently, Knight and Falke observed the dissociation of PH domains from supported bilayers followed by rapid rebinding to the surface 
after a short excursion in the bulk solution \cite{knight2009BJ}. 
They proposed that the hopping process may be important in the search for target molecules 
in eukaryotic cells. Subsequently, Yasui et al. found that PTEN (phosphatase
and tensin homologue) molecules hop along the plasma membrane of living
cells due to dissociation followed by rebinding \cite{Yasui2014}. 
PTEN is an important protein that suppresses development of cancer. It prevents cells from growing and dividing too rapidly
by dephosphorylating phosphoinositide substrates on the plasma membrane. 
PTEN-membrane affinity is regulated by a C2 domain and it is enhanced by electrostatic interactions. 
The observed hopping of the C2 domain on the plasma membrane is thus expected to alter the dynamics of the search for a phospholipid substrate. 

A straightforward consequence of membrane hopping is that a molecule remains in its immediate vicinity for a short time and 
then jumps to a location that is further away than expected from two-dimensional diffusion. Therefore, the search process is allowed 
to explore larger areas and the molecule can bypass diffusion barriers that may be present in the membrane. However, hopping comes at the cost of the search being less exhaustive. 
We may ask the questions how the dynamics of 
membrane-targeting domains is affected by such long jumps and how this motion deviates from a simpler two-dimensional 
diffusion. Such potential complex behaviour can yield anomalous diffusion of membrane-targeting domains, which would 
alter the outcome of search processes and the sequential molecular reactions.    

Anomalous diffusion is widespread in the motion of molecules in biological systems 
\cite{barkai2012PhysToday,hofling2013review,metzler2014review,krapf2015reviewCTM}. 
In general, a particle exhibits anomalous diffusion when the mean square displacement (MSD) scales as a power law 
with an exponent $\alpha \neq 1$
\begin{equation}
\langle x^2(t) \rangle=K_{\alpha} t^{\alpha}, \label{anomalous}
\end{equation}
where $K_{\alpha}$ is the generalized diffusion coefficient with units $\mathrm{cm}^2/\mathrm{s}^{\alpha}$. 
When $\alpha<1$ the process is subdiffusive and when $\alpha>1$ it is superdiffusive.
Subdiffusion in the cytoplasm \cite{goldingCox2006,tolic2004anomalous,jeon2011vivo}, 
the nucleus \cite{bronstein2009PRL}, 
and the plasma membrane \cite{weigel2013pnas,heinemann2013lateral,torreno2014enhanced} of live cells 
is caused by crowding \cite{banksFradin2005,szymanskiWeiss2009} 
and complex interactions with the cytoskeleton and macromolecular complexes, among others.
Similarly, subdiffusion can take place in model membranes due to crowding and packing effects  
\cite{horton2010development,jeon2012anomalous}. 
The appearance of superdiffusion processes in biomolecular systems is far less common. 
The archetypal mode of superdiffusive motion is due to active cytoplasmic flows and 
transport mediated by molecular motors, requiring ATP energy consumption 
\cite{bursac2007cytoskeleton,Kahana2008,bruno2009transition}. 

From a theoretical point, there are three major mechanisms that can introduce superdiffusion \cite{akimoto2012distributional}. 
It can be caused by correlations in the random walk, such as those in fractional Brownian motion with 
a Hurst index $H>1/2$, by persistent directional motions (L\'{e}vy walks), and by long jumps 
(L\'{e}vy flights). Active biological transport can be modelled as L\'{e}vy walks \cite{bruno2009transition}.
Bulk-mediated diffusion processes, which can be described as L\'{e}vy flights, have 
been observed for transient adsorption on a solid surface 
where molecules display intermittent behaviour, alternating between periods of 
immobilization at the solid-liquid interface and periods of diffusion in the 
bulk fluid \cite{skaug2013intermittent,yu2013single}.

\begin{figure}
\centerline{\includegraphics[width=\linewidth]{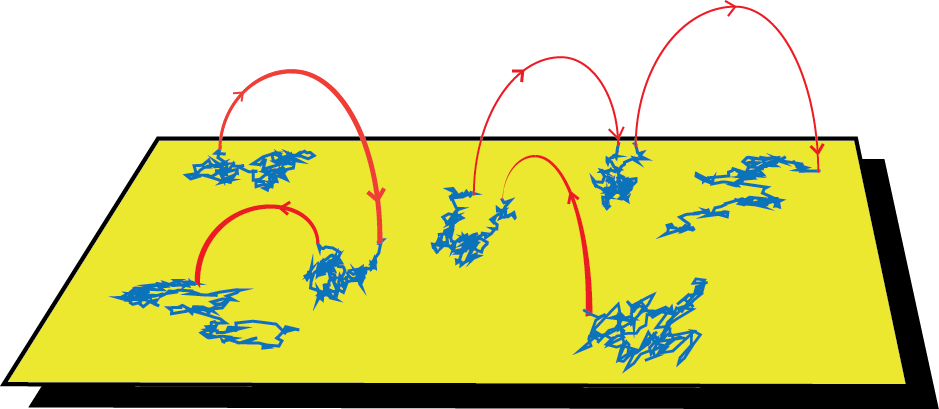}}
\vspace*{.05in}
\caption{\label{fig:sketch}Sketch of the diffusion process. A molecule alternates 
between phases of two-dimensional and three-dimensional diffusion. Diffusion in the three-dimensional 
bulk is much faster than diffusion on the lipid bilayer, and thus only the effective two-dimensional process 
is observed without loss of trajectory connectivity. The excursions into the bulk are seen as long jumps in the 
two-dimensional trajectories.}
\end{figure}

In this article we report the experimental observation of superdiffusive transport of membrane-targeting C2 domains 
on supported lipid bilayers. 
Measurements of the diffusion of membrane-targeting domains are performed by 
single-particle tracking and are compared to both analytical theory and numerical simulations. 
In stark contrast to active cytoplasmic transport, superdiffusion in model membranes 
does not require energy. Our data strongly suggests that superdiffusion is caused by bulk-mediated diffusion, namely 
molecules dissociate from the membrane and perform three-dimensional random walks until they reach the membrane 
again and readsorb at a new location, as sketched in Figure~\ref{fig:sketch}.  
Interestingly, the motion of membrane-targeting domains shows weak ergodicity breaking, 
a phenomenon that has recently attracted considerable attention in cellular environments and other complex systems 
\cite{bouchaud1992weak,bel2005weak,barkai2012PhysToday,krapf2013nonergodicity,metzler2014review}. 
The ergodic hypothesis, which is fundamental to statistical mechanics, states that 
ensemble averages and long-time averages of individual trajectories are equivalent. 
The violation of ergodicity has pronounced implications for the dynamics of individual molecules, 
which can be very different from the ensemble statistics \cite{barkai2012PhysToday}. 
In the traditional way of obtaining the MSD, 
the square displacements are averaged over a large ensemble of molecules at a time $t$ since the beginning of the measurement, 
i.e. an ensemble average. Alternatively, 
it is possible to perform the average over all the displacements in a lag time $\Delta$ of a single trajectory, i.e. a temporal average.
For ergodic systems, both averages converge to the same value. 
However, weak ergodicity breaking can take place as a consequence of kinetics with power-law statistics 
in the plasma membrane \cite{weigel2011PNAS,manzo2015PRX} 
and in the cytoplasm of live cells \cite{jeon2011vivo,tabei2013intracellular} as well 
as in inorganic complex systems such as quantum dots \cite{brokmann2003statistical,stefani2009beyond} 
and models of glassy dynamics \cite{bouchaud1992weak}.

\section*{Results}

\subsection*{Diffusion of membrane targeting proteins on supported lipid bilayers}

We tracked the motion of the membrane-targeting C2A domain from synaptotagmin 7 \cite{sugita2002synaptotagmins}, 
labelled with Atto-565, 
in a supported lipid bilayer composed of phosphatidylcholine (PC) and phosphatidylserine (PS) at a 3:1 ratio. 
The lipid bilayer was self-assembled on 
a clean cover\-slip \cite{knight2009BJ}. Imaging was done in a home-built total internal reflection (TIRF) microscope under continuous illumination at 20 frames/s. 
Surface densities were kept low enough to enable accurate 
tracing of trajectories and to allow assignment of connections even after micrometer-long jumps. 

\begin{figure}
%\centering
\centerline{\includegraphics[width=\linewidth]{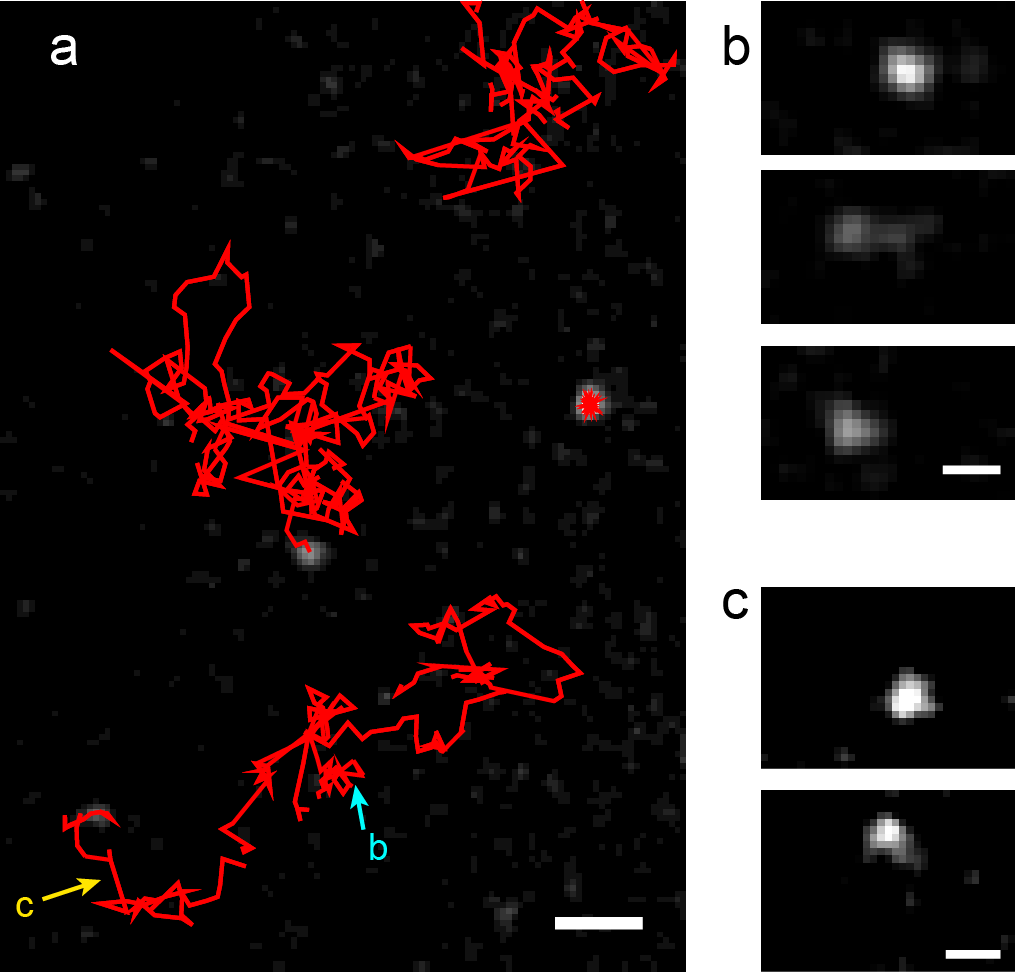}}
\vspace*{.05in}
\caption{\label{fig:tracks} 
Single particle tracking of membrane-targeting domains. 
(a) C2A-Atto565 individual trajectories collected during a 10-s time window. Three mobile trajectories are observed in the image 
together with one immobile particle that is tracked but is not included in the analysis. The last frame is overlaid  
on the trajectories. The original data is shown in 
Supplementary Video S1. Scale bar 2 $\mu \mathrm{m}$. 
(b) Region of interest (ROI) around the location of a micrometer jump that occur in the lowermost 
trajectory, marked with the letter b. Three frames are shown corresponding to before, during, and after the jump. 
Scale bar 0.5 $\mu \mathrm{m}$. 
(c) ROI around the location of the jump marked with the letter c. Scale bar 1 $\mu \mathrm{m}$.}
\end{figure}

Figure \ref{fig:tracks}a shows an example of trajectories obtained in a 10-s window, overlaid on the last frame. 
Often, long jumps are observed in the particle trajectories as seen in the examples in Figs. \ref{fig:tracks}b and c. These 
jumps suggest the C2A molecules detach from the surface and readsorb after brief excursions into the liquid bulk. 
The motion in the bulk is much faster than diffusion on the viscous membrane and jumps are thus expected to occur 
instantaneously for all practical purposes. For the C2A domain, the diffusion coefficient in the lipid bilayer $D_s$ is 
of the order of 2 $\mu \mathrm{m}^2/\mathrm{s}$, 
but in liquid the diffusion coefficient $D_b$ is estimated to be 100 times higher \cite{ziemba2013lateral}. 
As a consequence, when a molecule performs a jump through the bulk 
it can sometimes be observed at reduced intensity in both the old and new locations within the same imaging frame, 
as seen in Figure~\ref{fig:tracks}b. 

\begin{figure}
%\centering
\centerline{\includegraphics[width=\linewidth]{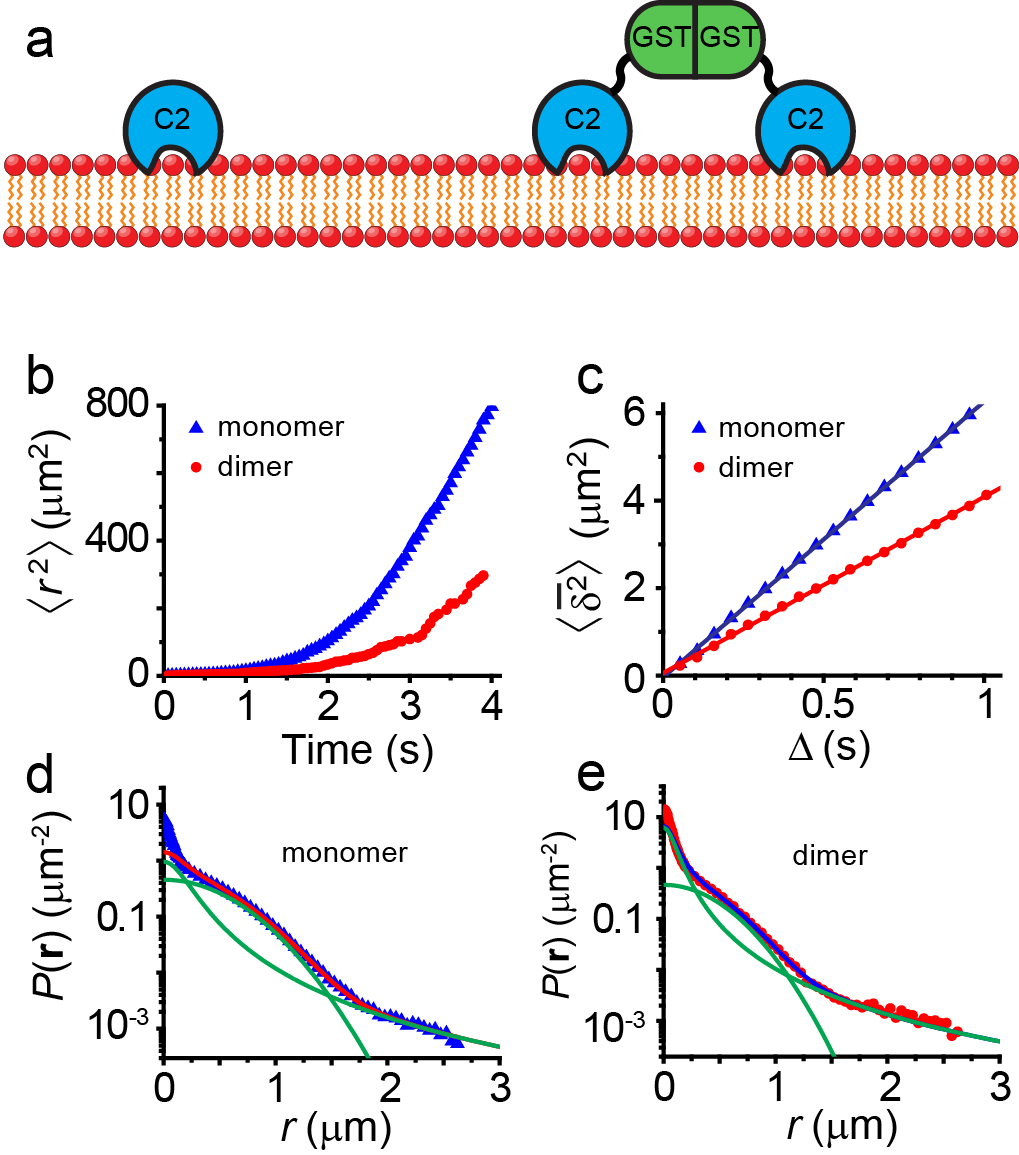}}
\vspace*{.05in}
\caption{\label{fig:C2spt} 
Anomalous diffusion analysis of membrane-targeting domain C2A (monomer) and dimer forming GST-C2A. 
(a) Sketch of the C2A monomer and the GST-C2A dimer employed in this study.
(b) Ensemble averaged MSD $\langle r^2(t) \rangle$. 
(c) Time averaged MSD $\langle \overline{\delta^2} \rangle$ as a function of lag time $\Delta$. 
The time average-MSD of individual trajectories varies greatly, so the MSDs of individual trajectories 
are also ensemble averaged.
(d-e) Distribution of displacements for $\Delta=100$ ms. 
The total number of displacements are 207,000 and 56,000 for C2A and GST-C2A, respectively. 
The solid lines show fitting to equation~(\ref{Levy}) and to the 
individual components of the propagator, i.e. the Gaussian part $[(1-\omega)/2\pi \sigma^2] \mathrm{exp}(-r^2/2\sigma^2)$
and the Cauchy propagator part $\omega \gamma/2\pi (r^2+\gamma^2)^{3/2}$.
The cutoff at 2.6 $\mu \mathrm{m}$ appears because trajectories are not connected when jumps longer than 
this distance take place. This threshold is placed in order to avoid the risk of particle misconnections.}
\end{figure}

In order to study the effect of the dissociation constant, we also employed a C2A construct fused to a non-membrane interacting 
glutathione S-transferase (GST), which has a strong tendency to dimerize (Figure~\ref{fig:C2spt}a).  
The GST-C2A dimer forms two independent interactions with the membrane and will consequently have a slower dissociation rate than 
C2A monomer, providing a good comparison for validating our superdiffusion predictions.
Additionally, GST-C2A dimer has a higher viscous drag coefficient and, in turn, its diffusion coefficient on the membrane surface is reduced 
to nearly half \cite{knight2010BJ}.  

We collected 14,000 C2A and 3,600 GST-C2A mobile trajectories. Immobile fluorophores that did not exhibit any apparent diffusive motion 
were excluded from the analysis. The ensemble-averaged MSD 
$\langle r^2(t) \rangle$ of C2A monomers and dimer-forming GST-C2A are shown in Figure~\ref{fig:C2spt}b. A deviation 
from a linear MSD is evident in the figure, showing superdiffusive behaviour. Further, the onset of superdiffusion for GST-C2A 
occurs at a later stage. 

The time-averaged MSD $\overline{\delta^2(\Delta)}$ is often used in the analysis of individual trajectories. 
Throughout this manuscript we will denote the ensemble average of an observable with brackets $\langle \cdot \rangle$ 
and the time average with an overbar $\overline \cdot$ .
For a trajectory with $N$ time points,  
\begin{equation}
\overline{\delta^2(\Delta)}=\frac{1}{N-n} \sum_{j=1}^{N-n} [\textbf{r}(j\tau+\Delta)-\textbf{r}(j\tau)]^2, 
\label{tMSDdef}
\end{equation}
where $\tau$ is the time interval between consecutive measurements and $n=\Delta/\tau$. 
This approach is especially useful when a limited number of trajectories is available, as usually occurs
in single-molecule studies. 
Figure~\ref{fig:C2spt}c shows the time-averaged MSD after it is additionally averaged 
over all the trajectories. 
GST-C2A exhibits the expected slower diffusion rate than C2A, based on the MSD slope.
As mentioned above, for ergodic processes, the temporal and ensemble averages coincide in the 
long time limit, $\overline{\delta^2(\Delta)}=\langle r^2(\Delta) \rangle$. 
However, the ergodic hypothesis breaks down for C2A molecules. 
In contrast to the ensemble-averaged MSD, the time-averaged MSD is linear in lag time 
\begin{equation}
\langle \overline{\delta^2(\Delta)} \rangle \sim \Delta. \label{tMSD}
\end{equation}
Thus, an observer analysing time-averages would reach the misleading conclusion 
that the diffusion behaviour is not anomalous. 
    
The distribution of displacements $P(\textbf{r})$ at $\Delta=100 \ \mathrm{ms}$ 
is shown in Figs. \ref{fig:C2spt}d and e for C2A and GST-C2A, respectively.  
The distribution exhibits two different characteristic 
regimes: a central part up to a distance $r\approx1.5 \ \mu\textrm{m}$ and a long tail. 
This behaviour can be understood from the scaling properties of bulk-mediated diffusion as discussed by 
Bychuk and O'Shaughnessy \cite{Bychuk1995}. Once a molecule dissociates from the surface, 
it performs a three-dimensional random walk until it returns. In the asymptotic limit, the first return time 
distribution scales as $\psi(\tau)\sim \tau^{-1.5}$. For any given 
return time, the surface distance between the dissociation and return points has 
a Gaussian distribution $P(\textbf{r}_j|\tau)\sim \mathrm{exp}(-r_j^2/4D_b \tau)$. 
Therefore, the distribution of jump lengths is $P(\textbf{r}_j)\sim r_j^{-3}$, 
as observed in Figs.~\ref{fig:C2spt}d and e for long distances.   

The theoretical probability density function of jump lengths can be found using the image method \cite{redner}. 
The distance of first return to the surface are governed by $P(\textbf{r})=\gamma_0/2\pi (r^2+\gamma_0^2)^{3/2}$, 
that is a two-dimensional Cauchy distribution. At short times, the probability that the particle performs 
more than a single jump is small. If we neglect the distance covered by surface diffusion 
within time intervals at which the particle undergoes a bulk excursion,  
the motion at each short interval is either by surface diffusion or via a jump. 
We can then approximate the distribution of displacements 
at short times by 
\begin{equation}
P(\textbf{r})=\omega \frac{\gamma_0}{2\pi (r^2+\gamma_0^2)^{3/2}} + \frac{(1-\omega)}{2\pi \sigma^2} \mathrm{exp}(-r^2/2\sigma^2), \label{Levy}
\end{equation}  
where $\omega$ is the probability that the particle hops within the given time
and surface diffusion yields $\sigma^2=2D_s t$.
A least-square fitting of the distribution of displacements (Figs.~\ref{fig:C2spt}d and e) to this propagator yields 
$D_s=1.7 \ \mu \mathrm{m}^2/\mathrm{s}$ for C2A monomers and $D_s=1.0 \ \mu \mathrm{m}^2/\mathrm{s}$ for GST-C2A. 
The parameter $\gamma$ is found to be 0.24 $\mu \mathrm{m}$ and 0.12 $\mu \mathrm{m}$ for C2A and GST-C2A, respectively. 

The distribution of displacements for longer times involves both 
a random number of jumps, each having a Cauchy distribution, and the Brownian motion on the surface. 
Chechkin et al. derived the full solution for the propagator of bulk-mediated diffusion \cite{Chechkin2012}. 
For the case when $D_s=0$ and neglecting long distance 
corrections, the distribution of displacements is given by the Cauchy propagator, 
in agreement with scaling arguments \cite{Bychuk1995}, 
\begin{equation}
P(\textbf{r})=\frac{\gamma t}{2\pi [r^2+\left(\gamma t \right)^2]^{3/2}}. \label{Cauchy}
\end{equation} 
When the particles also diffuse on the surface, i.e. $D_s\neq 0$ the 
probability density of the displacements is given by the convolution 
of equation~(\ref{Cauchy}) with a normal distribution. 
Even though the full solution for long times is complicated, the tail of this distribution for 
large distances still scales as $P(\textbf{r})\sim r^{-3}$.
Due to this asymptotic behaviour, the exact distribution 
has similar properties to the Cauchy distribution.

\subsection*{Numerical simulations: diffusion in the presence of bulk excursions}

In order to verify the model of surface diffusion in the presence of bulk excursions we 
analyse numerical simulations of the process diagrammed in Figure~\ref{fig:sketch}. 
Molecules perform a two-dimensional random walk, but at random times they jump 
due to a hypothetical bulk excursion. The surface residence times are assumed to be independent 
and identically distributed exponential random variables and the jumps are modelled according to 
the first return time to the surface given simple diffusion in a three-dimensional medium. 
These simulations are analysed in the same way 
as with experimental observations of the motion of membrane-targeting C2 
domains on supported membranes. 500 realizations were 
simulated off-lattice with a surface diffusion coefficient $D_s=0.5$ and a dissociation coefficient $k=0.1$. 
The chosen parameters do not intend to
capture the real protein properties, but to simply test theoretical predictions 
without the effects of experimental noise. 
The displacements for two-dimensional diffusion are drawn from a Gaussian distribution 
with variance $\sigma_s^2=1$ and the return times from bulk excursions are drawn from a distribution 
$\psi(t_b)=z_0 (4\pi D_b t_b^3)^{-1/2} \mathrm{exp}(-z_0^2/4D_b t_b)$ \cite{redner}. 
Then the jump distances are drawn from a Gaussian distribution with variance $\sigma_b^2=2D_b t_b$.

\begin{figure}
%\centering
\centerline{\includegraphics[width=\linewidth]{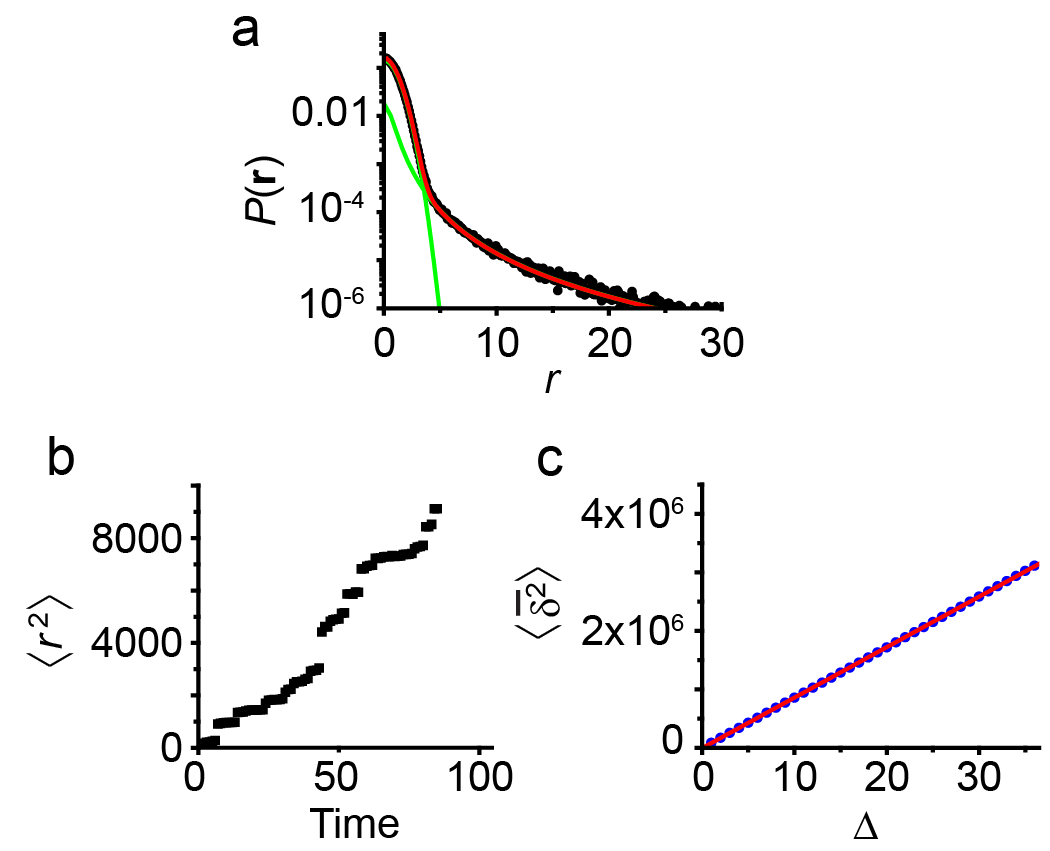}}
\vspace*{.05in}
\caption{\label{fig:simulations} 
Numerical simulations of L\'{e}vy flights. 
500 realizations were performed, in which a particle alternates between 2D random walks and bulk-mediated jumps. 
(a) Probability density of the tracer displacements. The density is well described by a theoretical model 
that includes a Gaussian central part and a Cauchy propagator of the form $\gamma_0/(r^2+\gamma_0^2)^{3/2}$.
(b) Ensemble-averaged MSD  $\langle r^2(t) \rangle$ as a function of time. The ensemble-averaged MSD is computed 
from the distance covered by the 
tracer in a time $t$ from the start of the realization.
(c) The time-averaged MSD $\overline{\delta^2(\Delta)}$ is averaged over all realizations and plot against lag time 
$\Delta$.}
\end{figure}

The distribution of displacements $P(\textbf{r})$ 
for the numerical simulations is shown in Figure~\ref{fig:simulations}a. 
As expected, there are two regimes: a central Gaussian part due to the two-dimensional diffusion 
on the membrane between bulk excursions, and a heavy tail that 
arises from the long distance behaviour of bulk excursions. 
The distributions for short times can again be modelled with a 
propagator that includes contributions from Gaussian surface diffusion and a Cauchy distribution 
due to bulk excursions. 
By fitting to equation~(\ref{Levy}), it is found $D_s=0.50\pm0.05$ 
(the value employed in the simulations is $D_s=0.5$) and $\gamma_0=0.75$.

\subsection*{MSD analysis}

The dynamics of a particle with a Cauchy propagator are 
particularly interesting because the theoretical variance of the displacements diverges,
\begin{equation}
\langle r^2(t) \rangle = \int_0^{\infty} (2\pi r) r^2 P(\textbf{r}) dr = \infty. \label{variance}
\end{equation}
In practice, a diverging second moment implies that there is a non-negligible probability for the occurrence 
of extremely long jumps and this phenomenon has direct implications in the measured MSD.
Figure \ref{fig:simulations}b shows the ensemble-averaged MSD computed from the numerical simulations. 
The MSD increases in a superlinear fashion, i.e. by employing 
equation~(\ref{anomalous}), we have $\alpha>1$, which implies the process is superdiffusive. 

%%%%%%%%%%%%%%%%%%%%%%%%%%%%%%

Let us now analyse the unexpected MSD behaviour, starting from the time-averaged MSD 
of individual trajectories.  We can show that the time-averaged MSD is linear in lag time 
for any random walk with independent increments $\textbf{u}_i=\textbf{r}_{i+1}-\textbf{r}_i$, such that 
$\langle \textbf{u}_i \cdot \textbf{u}_j \rangle=0$ when $i \neq j$. From the definition of the time 
averaged MSD (equation~(\ref{tMSDdef})) \cite{froemberg2013random},
\begin{eqnarray}
\overline{\delta^2(\Delta)} 
& = &\frac{1}{N-n}\sum_{i=0}^{N-n} \left(\sum_{k=i}^{i+n}\textbf{u}_k\right)^2 \\
& \approx & \frac{1}{N-n}\sum_{i=0}^{N-n} \sum_{k=i}^{i+n}\textbf{u}_k^2 \\
& \approx & \frac{\Delta}{t} \sum_{i=1}^{N-n} \textbf{u}_i^2, \label{tMSD-approx}
\end{eqnarray}
where we have used the approximation that $t\gg\Delta$, we omitted the term 
$\sum_i\sum_{j \neq i}\textbf{u}_i \cdot \textbf{u}_j$ because it is zero on average, 
and again we have used the parameter $n=\Delta/\tau$. 
Therefore we see that for symmetric random walks with independent increments, 
the time-averaged MSD is linear as observed in Figs.~\ref{fig:C2spt}c and \ref{fig:simulations}c.

Although the time-averaged MSD for individual trajectories is linear, 
the ensemble averaged MSD $\langle r^2(t) \rangle$ is not. We can understand 
the superdiffusive behaviour by assuming we can define the motion in terms 
of two independent processes $\textbf{r}(t)=\textbf{b}(t)+\textbf{y}(t)$, where 
$\textbf{b}(t)$ is a two-dimensional Brownian motion and $\textbf{y}(t)$ is 
a L\'{e}vy process with a probability density defined by equation~(\ref{Cauchy}). Then the MSD is 
$\langle \textbf{r}^2 \rangle =\langle \textbf{b}^2 \rangle +\langle \textbf{y}^2 \rangle $. 
The first term is linear in time but the second term has a superdiffusive nature 
\cite{Bychuk1995,valiullin1997levy,Chechkin2012}.

\section*{Discussion}

The propagator for surface diffusion in the presence of bulk-mediated jumps (equation~(\ref{Levy})) depends on the 
surface diffusion coefficient $D_s$ 
and the parameter $\gamma$ that reflects the transition between the surface and the bulk phase. 
Namely, $\gamma \sim a/\tau_{\mathrm{des}}$, where $\tau_{\mathrm{des}}$ is the mean desorption time 
and $a$ is a dimensional factor. 
Bulk-mediated diffusion thus predicts $\gamma_\mathrm{dimer}<\gamma_\mathrm{monomer}$, 
in agreement with the values we find for C2A and GST-C2A. 

The surface motion of these membrane-targeting domains is well described by L\'{e}vy flights, 
a random walk where the step displacements have a heavy-tailed distribution. 
The heavy tail arises from the dissociation of molecules from the membrane, which then perform a three-dimensional 
random walk until they reach the surface again at another location. The 
process involves the first return to a surface and it converges to a power law according to the Sparre-Andersen 
theorem \cite{redner}. 
This type of L\'{e}vy flight dynamics is fundamentally different from 
L\'{e}vy walks induced by molecular motors in the cytoplasm because periods of active motion require an energy input, 
typically in the form of ATP hydrolysis, while bulk excursions occur spontaneously.

One of the most interesting effects of the observed bulk-mediated diffusion statistics 
is that the ensemble-averaged MSD exhibit superdiffusive behaviour, whereas the temporal averages  
suggest normal diffusion. This nonergodic behaviour is similar to that of continuous time 
random walks (CTRW) where the sojourn time distribution between steps  
has a probability distribution that is heavy-tailed. Also in the CTRW, $\langle \overline{\delta^2(\Delta)} \rangle\sim \Delta$ 
and $\langle r^2(t) \rangle\sim t^\alpha$, albeit the CTRW is subdiffusive with $\alpha<1$. 
The difference in the behaviour of temporal and ensemble averages is the key signature of weak ergodicity breaking in 
the process \cite{margolin2006nonergodicity}. 

To date, different groups have observed normal diffusion for membrane proteins in supported lipid bilayers, which 
appear to contradict our findings \cite{tamm1988lateral,gambin2006lateral,ramadurai2010influence,ziemba2012assembly}. 
There are several reasons for this apparent discrepancy. 
Single-particle tracking in lipid bilayers often focuses on time-averaged MSD, which does not show 
any non-linearity in lag time. Thus it would be reasonable to reach the conclusion that diffusion is not anomalous. 
Furthermore, anomalous diffusion in supported bilayers is known to develop as a result of 
packing and crowding. These mechanisms are modelled by a fractional Langevin equation, which is ergodic in nature, with anomalies 
that show up in the time averages. The distribution of displacements has also been previously reported 
as exhibiting Gaussian behaviour. Here we report on the motion of surface-bound membrane domains that 
exhibit desorption from the membrane within the experimental observation time. The behaviour  
of transmembrane proteins or lipids is very different because the free energy barrier for release from the membrane is 
too high to be observed within the constrains of experimental observations 
\cite{tamm1988lateral,gambin2006lateral,ramadurai2010influence}. Previous works dealing 
with membrane-targeting domains such as C2 have generally been limited to short displacements in order to exclude the effect 
of long bulk-mediated jumps in diffusion measurements \cite{knight2010BJ,ziemba2012assembly}.

What are the biological implications of surface superdiffusion for peripheral membrane proteins? 
Search processes are ubiquitous in cell biology and it is feasible to assume that evolution has 
optimized search parameters. For signalling molecules delivered to the plasma membrane during a specific stimulus, 
the target molecule is often scarce in a sea of other lipids and proteins. Thus we can envision 
that if a molecule does not find its target in a given time, it becomes more efficient to start searching at 
a different location. Is it appropriate then to assume L\'{e}vy flights yield the 
optimal search for sparse targets when compared to Brownian motion?  
For one-dimensional intermittent processes 
that switch between Brownian motion and ballistic relocation phases, 
it has been shown that the search process is significantly more efficient 
when relocation times are power-law distributed, resulting in a L\'{e}vy walk \cite{lomholt2008levy}. 
Notably, when L\'{e}vy dynamics are employed, the search is less sensitive to the target density \cite{lomholt2008levy}.
In general, the optimal strategy depends on the average target distance from the starting point \cite{Palyulin25022014}. 
However, blind searches inside a living cell are very different from a search in an unobstructed environment. 
Several aspects provide additional complexities in the plasma membrane, in particular \cite{krapf2015reviewCTM}. 
Experimental measurements show that the plasma membrane is compartmentalized in a way that 
proteins and lipids have the tendency to remain transiently confined within small regions \cite{ritchie2003fence}. 
Further, membrane proteins typically exhibit subdiffusion with 
anti-persistent increments where molecules drift towards the locations that 
they visited in the past. While this subdiffusive behaviour provides the opportunity for a thorough and compact search, it is definitely 
not the optimum situation to find sparse targets. A superdiffusive L{\'e}vy flight provides a 
mechanism to overcome the effects of anti-persistent correlated subdiffusive motion. Thus, we 
expect L{\'e}vy flight dynamics to often outperform a Brownian search.  

The obstruction to the diffusion of membrane molecules has two different sources, both of them causing 
anti-persistent correlations in the random walk. On one hand, obstacles can be introduced by immobile transmembrane proteins 
which affect all lipids and membrane proteins.  On the other hand, a more severe obstruction can be caused by cytoskeleton 
components that may not be in direct contact with the plasma membrane \cite{andrews2008actin}. 
The effect of these barriers is not equal for all membrane proteins. 
Proteins that have large intracellular complexes are blocked much more efficiently than small molecules. In cases where 
a large signalling molecule adheres to the membrane via phospholipid-binding domains, bulk excursions allow 
for the exploration of larger areas. Otherwise, the molecule would remain confined for long times within cytoskeleton-formed 
corrals, even when no substrate target is found within this region.   

In summary, we have observed the non\-ergodic, superdiffusive motion of membrane-targeting peptide domains in supported lipid 
bilayers. The motion is well-described by L\'{e}vy flights with jumps that have a heavy-tail distribution. 
The long jumps are caused by excursions 
into the liquid bulk. After dissociating from the membrane, the molecules diffuse in three dimensions until they 
reach the membrane again and bind at a new location. Diffusion in the liquid bulk is much faster than diffusion in 
the membrane, therefore we do not consider the delay time between dissociation and readsorption. 
The surface distances covered by jumps have a Cauchy distribution, which is responsible
for the heavy tail in the superdiffusive L\'{e}vy flights. 
Model membranes provide an elegant system to study the effect of superdiffusive L\'{e}vy flights 
because they are not subjected to the interactions with other cell components that would mask its experimental observation. 
However, hopping was already observed on the surface of live cells \cite{Yasui2014} and 
we foresee these processes have broad physiological relevance in the surface diffusion of signalling molecules. 

\section*{Methods}

\subsection*{Imaging buffer}
Imaging and rinsing during the preparation steps was performed in an imaging buffer consisting of 
50 mM HEPES, 75 mM NaCl, 1 mM MgCl$_2$, 2 mM tris(2-carboxyethyl)phosphine (TCEP), 200 $\mu$M CaCl$_2$.
CaCl$_2$ is necessary for C2 domain binding to the reconstituted membrane.

\subsection*{Preparation of phospholipid vesicles}
Phospholipids were purchased from Avanti Polar Lipids (Alabaster, AL). 
Chloroform-suspended 18:1 ($\Delta 9$-Cis) PC (DOPC) and 18:1 PS (DOPS) were mixed at a ratio of 3:1. 
The phospholipid mixture was vacuum dried overnight and 
resuspended in imaging buffer to a final concentration of 3 mM followed by probe sonication 
to form sonicated unilamellar vesicles (SUVs). 

\subsection*{Preparation of coveslips and supported lipid bilayers}
Glass cover\-slips were cleaned by sonication in a detergent solution followed by soaking in 1M KOH. 
The cover\-slips were rinsed extensively in Milli-Q water and blown dry with a stream of nitrogen gas. 
Then, the cover\-slips were treated with an oxygen plasma. Immediately after the plasma cleaning, 
a perfusion chamber (CoverWell, Grace Bio-Labs) was adhered to the coverslip. 
In order to deposit the lipid bilayers a solution of SUVs (1.5-mM lipid) 
composed of phosphatidyl\-choline (PC) and phosphatidyl\-serine (PS) at a 3:1 ratio in 1M NaCl and imaging buffer 
was introduced into the perfusion chamber 
and incubated for one hour at 4$^{\circ}$C. 
Refrigeration minimizes lipid oxidation. 
The surface was then rinsed with imaging buffer multiple times prior to addition of protein sample. 

\subsection*{C2A and GST-C2A expression and purification}
An expression plasmid containing the gene for a GST-ybbR-Synaptotagmin 7 (Syt7) C2A domain fusion protein was transformed into 
{\it E. coli} BL21-CodonPlus(DE3) competent cells. The ybbR segment provides a site for Sfp-catalysed fluorophore labelling \cite{ybbr}. 
Cells were grown at 37$^{\circ}$C to an OD$_600$ of 0.6 
and then induced to express protein with 0.5 mM IPTG at room temperature for 6 hours. 
The harvested cells were lysed at 18,000 lb/in$^2$ in a microfluidizer 
in a buffer containing 50 mM Tris pH 7.5, 400 mM NaCl and centrifuged at 17,000 rpm in a Sorval SS-34 rotor. 
The clarified lysate was loaded onto a 5-ml GSTrap FF column (GE Healthcare LifeSciences, Pittsburgh, PA) 
followed by gradient elution with 50 mM Tris, pH 8.0, 100 mM NaCl, and 10 mM glutathione. 
Fractions containing protein were pooled and diluted to reduce the salt to less than 0.1 M prior 
to loading onto a HiTrap Q HP column (GE Healthcare LifeSciences, Pittsburgh, PA) and eluting 
with a linear gradient to 1 M NaCl in 25 mM Tris, pH 8.5, 20$\%$(vol/vol) glycerol, and 0.02$\%$(wt/vol) NaN$_3$. 
A portion of the construct was subjected to thrombin cleavage and then separated using a 
Superdex 200 gel filtration column (GE Healthcare LifeSciences, Pittsburgh, PA) 
equilibrated in 50 mM Tris, pH 7.5 and 100mM NaCl to yield a ybbr-Syt7 C2A construct.  

\subsection*{Protein labeling} 
10 mM CoASH (New England Biolabs, Ipswich, MA) in 400 mM Tris, pH 7.5 was mixed with 10 mM ATTO-565 maleimide (ATTO-TEC, Siegen, Germany) 
in dimethyl\-formamide and incubated at 30$^{\circ}$C overnight to form ATTO-565 CoA, then quenched with 5 mM DTT, 10 mM Tris pH 7.5. 
10 $\mu$M GST-ybbr-Syt7 C2A and ybbr-Syt7 C2A were labelled with the ATTO-565 via SFP synthase (4$\prime$-phosphopantetheinyl transferase). 
Each reaction contained 50 mM tris 7.5, 10 mM MgCl$_2$, 40 mM NaCl, 20 $\mu$M ATTO-565 CoA and 1 $\mu$M SFP synthase. 
Reactions were incubated at room temperature for 30 minutes, then placed at 4$^{\circ}$C overnight. 
Samples were dialysed against 1 L of 50 mM HEPES, pH 7.0, 75 mM NaCl, 4 mM MgCl$_2$ and 5$\%$ 
glycerol overnight at 4$^{\circ}$C then concentrated to 10 $\mu$M.

\subsection*{Imaging}
All images were acquired using an objective-type total internal reflection fluorescence
microscope (TIRFM). The microscope was home-built around an Olympus IX71 body 
\cite{weigel2011PNAS,weigel2013pnas} with a 561~nm laser line 
as excitation source. A back-illuminated electron-multiplied charge coupled device (EMCCD)
camera (Andor iXon DU-888) liquid-cooled to -85$^{\circ}$C, with an electronic gain of 300 was used.
In order to maintain constant focus during the whole imaging time we employed an autofocus system 
(CRISP, Applied Scientific Instrumentation, Eugene, OR) in combination with a piezoelectric stage 
(Z-100, Mad City Labs, Madison, WI). Videos were acquired at a frame rate of 20 frames/s.

\subsection*{Image processing and single-particle tracking}
Images were acquired using Andor IQ 2.3 software and saved as 16-bit tiff files. 
Then the images were filtered using a Gaussian kernel with a standard deviation of 1.0 pixel in ImageJ.
Single-particle tracking of Atto-C2 and Atto-GST-C2 was performed in MATLAB 
using the U-track algorithm developed by Jaqaman et al. \cite{jaqaman2008U-track} 
under thorough manual inspection of detection and tracking.

\bibliographystyle{pnas2011}
% \bibliography{Distbib}

\section*{Acknowledgments}
We thank Jeff Knight for kindly supplying the plasmids for C2A and C2A-GST and for useful discussions.
We also thank Kassi Prochazka for help in the initial part of the project and 
for her help in producing Figure~\ref{fig:sketch}.
This work was supported by the National Science 
Foundation under grant 1401432 (to DK) and by the National 
Institutes of Health under grant R21AI111588 (to OBP).

\section*{Author contributions statement}
G.C., O.B.P., and D.K. conceived the experiments; O.B.P. and D.K.supervised the project; 
G.C., K.N. and B.W.S. conducted the experiments; 
D.K. designed the analytical model and performed numerical simulations; 
K.N. and D.K. analysed the results; D.K. wrote the first draft; G.C., K.N., O.B.P.,and D.K. reviewed the manuscript. 

\section*{Competing financial interests}
The authors declare no competing financial interests.

\end{document}